\documentclass{emulateapj} 
\input{psfig.sty}
\usepackage{natbib}
\newcommand{\kms}{\,{\rm km\,s^{-1}}}
\newcommand{\msun}{\,{\rm M_\odot}}

\newcommand{\etal}{{et al.\ }}

\newcommand{\beq}{\begin{equation}}
\newcommand{\eeq}{\end{equation}}
\newcommand{\ba}{\begin{eqnarray}}
\newcommand{\ea}{\end{eqnarray}}
\def\spose#1{\hbox to 0pt{#1\hss}}
\newcommand{\lta}{\mathrel{\spose{\lower 3pt\hbox{$\mathchar"218$}}
      \raise 2.0pt\hbox{$\mathchar"13C$}}}
\newcommand{\gta}{\mathrel{\spose{\lower 3pt\hbox{$\mathchar"218$}}
      \raise 2.0pt\hbox{$\mathchar"13E$}}}
\def\simlt{\mathrel{\rlap{\lower 3pt\hbox{$\sim$}}\raise 2.0pt\hbox{$<$}}}
\def\simgt{\mathrel{\rlap{\lower 3pt\hbox{$\sim$}} \raise 2.0pt\hbox{$>$}}}
\lefthead{Volonteri \& Madau}
\righthead{Off-nuclear AGN}

\begin{document}
\submitted{}
\title{Off-nuclear AGN as a signature of recoiling massive black holes}

\author{Marta Volonteri\altaffilmark{1} \& Piero Madau\altaffilmark{2}}
\altaffiltext{1}{Astronomy Department, University of Michigan, Ann Arbor, MI, 48109.}
\altaffiltext{2}{Department of Astronomy \& Astrophysics, University of 
California, Santa Cruz, CA 95064.}

\begin{abstract}
During the final phases of inspiral, a massive black hole (MBH) binary experiences a 
recoil due to the asymmetric emission of gravitational waves. We use recent results 
from numerical relativity simulations together with models of the assembly and growth 
of MBHs in hierarchical cosmologies, to study the dynamics, statistics, and 
observability of recoling MBHs. We find that, at redshift $z<3$, kicked non-rotating 
holes are typically found between 1 and 30 kpc from their galaxy centers, while 
rapidly rotating ones are typically between 10 and a few hundred kpc. 
A recoiling hole that carries an accretion disk may shine as an ``off-nuclear AGN" 
while it moves away from the center of its host galaxy. We predict that, depending on 
the hole spin distribution and the duration of their active phase, a population of 
off-nuclear AGN may already be detectable at low and intermediate redshifts in present 
deep {\it Hubble Space Telescope} observations. The {\it James Webb Space Telescope} 
may discover tens of wandering AGN per square degree, most of them moving within their 
host halos on unbound trajectories.
\end{abstract}


\section{Introduction}

The massive black holes (MBHs) observed today at the centers of nearby galaxies \citep[see, 
e.g.,][]{richstone1998} are expected to have grown through a series of gas accretion 
episodes, when they shine as active galactic nuclei (AGN), and of binary coalescences
following the merger of their host galaxies \citep[e.g.,][]{BBR1980,VHM}. 
Coalescing MBH pairs will give origin to the loudest gravitational wave events in 
the universe, and are one of the primary targets for the planned {\it Laser 
Interferometer Space Antenna} ({\it LISA}; e.g. \citealt{Sesanaetal2004}). In general, 
gravitational waves also remove net linear momentum from the binary and 
impart a kick to the center of mass of the system. 
The outcome of this ``gravitational rocket'' has been the subject of 
many recent numerical relativity studies. Non-spinning holes 
recoil with velocities below 200 $\rm{km\,s^{-1}}$
that only depend on the binary mass ratio, whereareas much larger kicks are predicted
for rapidly rotating holes \citep{Campanelli2007,Gonzalez2007,Herrmann}. 

If it is not ejected from the host altogether (e.g. \citealt{Madau2004,Merrittetal2004,
Haiman2004,VolonteriRees2006,SchnittmanBuonanno2007}),
the recoiling MBH will travel some maximum distance and then return to the center
subject to dynamical friction \citep{MadauQuataert2004}. Historically, MBH ejections have been proposed  to explain the nature of radio jets \citep{saslaw,valtonen} and trails and filaments in spiral and interacting galaxies \citep[e.g.,][]{arp1976,arp2001}. \citet{haque} present a comprehensive compendium of early ideas on MBH ejections that have been recently revitalized. 

Galaxy mergers are frequent at early times, so a significant number of coalescing
MBH binaries is expected to form then. Galaxy mergers are also a 
leading mechanism for supplying gas to their nuclear black holes, and a recoiling hole 
can retain the inner parts of its accretion disk, providing fuel for a continuing luminous
phase along its trajectory. Two possible observational manifestations of 
gravitational-radiation ejection have then been suggested: (1) off-nuclear AGN activity 
\citep{MadauQuataert2004,Loeb2007}; and (2) broad emission lines that are substantially shifted    
in velocity relative to the narrow-line gas left behind \citep{Bonningetal2007}. 
In this Letter we use models of the assembly and growth of MBHs in hierarchical 
cosmologies to study the statistics and dynamics of recoiling holes, and explore the conditions
dictating their observability as off-nuclear AGN in deep optical and X-ray imaging studies. 

\section{Basic model}

We follow the assembly history of dark matter halos and associated MBHs via cosmological 
Monte Carlo realizations of the merger hierarchy from early times to the present in a
$\Lambda$CDM cosmology, using a model that captures many features of the 
MBH/AGN population (e.g., luminosity function of quasars and its evolution, black 
hole mass density today, stellar core formation due to MBH binary mergers). The main 
assumptions of the models have been discussed elsewhere 
(e.g. \citealt{VHM,VMH,Volonteri2007} and references therein). MBHs get incorporated through 
halo mergers into larger and larger structures, sink to the center owing to dynamical 
friction against the dark matter background, and form bound binaries. The MBHs 
in galaxies undergoing a major merger (i.e. having a mass ratio $>1:10$) have masses 
corresponding to the $M_{\rm BH}-\sigma_*$ relation of their hosts, i.e. at coalescence
the MBH binary mass ratio scales with the central velocity dispersions of the progenitor 
halos. After the formation of the binary, the black hole pair shrinks and coalesces on a 
timescale that is shorter than the merger time of the progenitor halos, as a result of 
stellar and gas-dynamical processes (e.g. \citealt{Bercziketal2006,Mayeretal2007}).

The recoil velocity $\vec{V}_{\rm kick}$ depends on the binary mass ratio $q=M_2/M_1$, on the 
dimensionless spin vectors of the pair $\vec{a}_1$ and $\vec{a}_2$ ($0<a_i<1$), and on 
the orbital parameters. All current numerical data on kicks can be fitted by \citep{Baker2008}
\begin{eqnarray}
\vec{V}_{\rm kick} &=& v_m \, \vec{e}_x + v_{\perp} (\cos\xi \, \vec{e}_x + \sin\xi \, \vec{e}_y) 
+ v_{\parallel} \, \vec{e}_z, \label{eq:v_total}\\
      v_m     &=& A \eta^2 \sqrt{1 - 4 \eta}\, (1 + B \eta), \label{eq:v_mass}\\
v_{\perp}     &=& H \eta^2(1+q)^{-1}\left( a_2^{\parallel} - q a_1^{\parallel} 
\right), \label{eq:v_perp}\\
v_{\parallel} &=& K \eta^3(1+q)^{-1}\,\cos(\Theta-\Theta_0)\left(a_2^{\perp} - q 
a_1^{\perp}\right), \label{eq:v_parallel}
\end{eqnarray}
where $\eta\equiv q/(1+q)^2$ is the symmetric mass ratio, the indices ${\parallel}$
and ${\perp}$ refer to projections parallel and perpendicular to the orbital angular momentum,
respectively, $\vec{e}_x$ and $\vec{e}_y$ are orthogonal unit vectors in the orbital
plane, $\Theta$ is the angle between $(\vec{a}_2^{\perp} - q {\vec a}_1^{\perp})$ and the separation
vector at coalescence, and $\Theta_0$ is some constant for a given mass ratio.  
Here, $A = 1.35 \times 10^4 \kms$, $B = -1.48$, $H = 7540 \pm 160 \kms$, $\xi =
215^\circ \pm 5^\circ$, and $K = 2.4 \pm 0.4 \times 10^5 \kms$ \citep{Baker2008}.
We will assume in the following that the orbital parameters and the spin orientations are 
isotropically distributed. The configuration producing the maximum recoil kick
(equal-mass rapidly rotating holes with anti-aligned spins oriented parallel to the 
orbital plane, with the pair recoiling along the $z$-axis), $V_{\rm kick}=K\eta^3=3750\,\kms$, is 
then quite rare (c.f. \citealt{Bogdanovic2007}).
\begin{figure}[thb]
\vspace{-0.3cm}
\includegraphics*[width=0.49\textwidth]{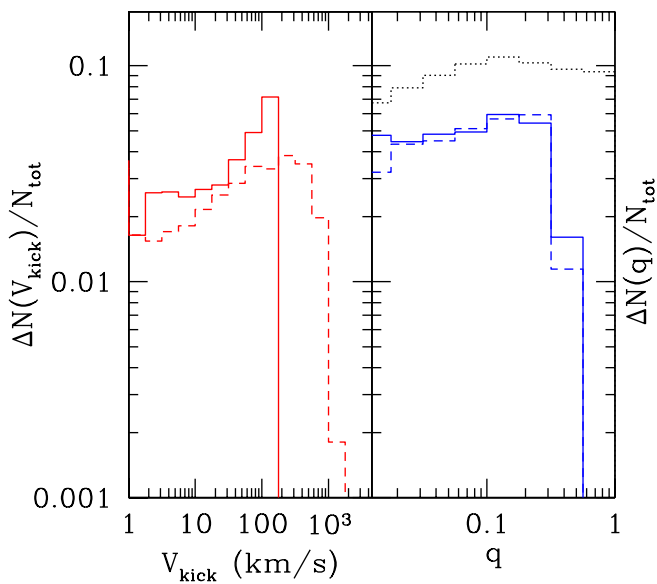}
\caption{\footnotesize {\it Left panel:} Kick velocity distribution for all $z<3$ MBH binaries 
with $M_1+M_2>10^5\,\msun$ in our cosmological realizations. {\it Solid curve}: nonspinning 
holes. {\it Dashed curve:} rapidly spinning holes with $a_i=0.9$. The orbital parameters and spin 
vectors are assumed to be isotropically distributed. {\it Right panel:} distribution of binary 
mass ratios. {\it Solid curve}: nonspinning 
holes. {\it Dashed curve:} rapidly spinning holes. In the latter case more holes are
ejected, and the distribution of binary mass ratios is slightly different. {\it Dotted curve:} Monte Carlo sampling 
the $z=0$ MBH mass function.} 
\label{fig1}
\end{figure}
Little is known about the masses of MBH binaries and their spins. The distribution of all 
binary mass ratios expected in hierarchical models is found to be relatively flat \citep{Volonterietal2005}. 
In this work we focus on binaries that could give origin to detectable off-center AGN: we 
therefore select pairs coalescing at $z<3$ with total masses $M_1+M_2>10^5\,\msun$. The 
distribution of mass ratios for this subsample is shown in Figure~\ref{fig1}. 
For comparison, we have also derived a distribution of mass ratios by Monte Carlo sampling 
the $z=0$ MBH mass function \citep[adopting the Schecter fit in][]{MerloniHeinz}, where 
we have assumed $M_1,M_2\in[10^4\msun,10^{10}\msun]$, and we have imposed the condition $M_1+M_2>10^5\,\msun$. 
In the sampling of the  $z=0$ mass function most binaries are composed of two rather small holes of similar mass, skewing the mass ratio towards unity ($\langle q\rangle=0.17\pm0.23$, implying $V_{\rm kick}\sim 250\kms$ for $a_i=0.9$).  The binary mass threshold we impose in the cosmological sampling 
also causes a loss of coalescing pairs with $q\approx 1$, because of the limited major merger 
activity of the corresponding host galaxies (of order of 0.1 Gyr$^{-1}$, \citealt{White07}). 

Large spins are a natural 
consequence of prolonged disk accretion episodes (e.g. \citealt{Bardeen1970,Volonterietal2005}), 
but can be avoided if AGN are fed by a series of small-scale, randomly oriented accretion 
events \citep{King2007}. 
To bracket the uncertainties, we consider here two extreme scenarios: prior to coalescence, 
both holes are assumed to be either non-rotating or are
rapidly spinning with $a=0.9$. The resulting distributions of kick velocities for 
these two cases are plotted in Figure \ref{fig1}.
Close to 7\% of all coalescing non-rotating holes are ejected from the nucleus with 
kick velocities between 100 and 200 $\kms$; in the spinning case, 12\% get kicks
above $100\,\kms$ and 2\% above $500\,\kms$.  

\section{Dynamics of recoiling holes}

In our default model the host galaxy is described by two spherical mass components:
a) the dark matter follows a \citet{NFW1997} profile 
with median concentration $c_{\rm med}=9(1+z)^{-1}(M_{\rm halo}/8\times 10^{12} 
h\msun)^{-0.14}$ and scatter around the median of $\Delta \log 
c=0.14$ \citep{Bullock2001}; 2) the central stellar profile is modeled instead 
as an isothermal sphere with one-dimensional velocity dispersion $\sigma_*$ and density
\beq
\rho_*(r)=\frac{\sigma_*^2}{2\pi G (r^2+R^2_{\rm BH})}. 
\eeq 
The velocity dispersion is related to the halo circular velocity $V_c$ at the 
virial radius by $V_c=\sqrt{2}\sigma_*$ (cf. \citealt{Ferrarese2002}). 
We truncate $\rho_*$ at the ``bulge'' radius $R_B=6\,$kpc $(\sigma_*/200\,\kms)^2$:
this ensures that the stellar bulge mass within $R_B$ is equal to $1000\,M_{\rm BH}$ 
\citep{HaringRix2004}, using $M_{\rm BH} = (10^8\,\msun)~(\sigma_*/200\,\kms)^4$ 
\citep{Tremaine2002}. The stellar profile is also truncated inside a core radius $R_{\rm BH}$, 
as stars within the gravitational sphere of influence of the hole, $R_{\rm BH}=GM_{\rm 
BH}/\sigma_*^2$, are bound to it and do not contribute to dynamical friction. 
The escape speed from such a 2-component potential can be approximated as 
$v_{\rm esc}\sim 2\sigma [c^2f(c)^{-1}(1+c)^{-1}+\ln(R_B/R_{BH})]^{1/2}$, where
$f(c)=\ln(1+c)-c/(1+c)$.
\begin{figure}[thb]
\vspace{-0.3cm}
\begin{center}
\includegraphics*[width=\columnwidth]{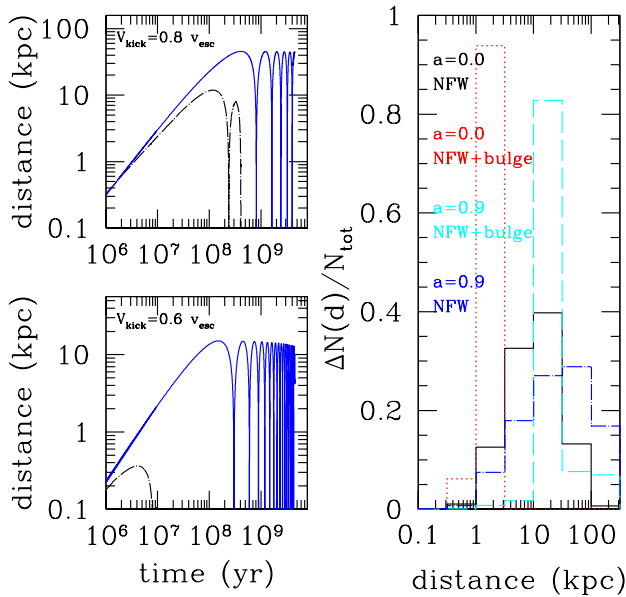}
\end{center}
\caption{\footnotesize {\it Left panels:} Decay of a $10^6\msun$ MBH in a spherical 
halo of mass $M_{\rm halo}=10^{12}\,\msun$ and escape velocity $v_{\rm esc}=380\,\kms$,
for $V_{\rm kick}=0.8\,v_{\rm esc}$ ({\it top}) and $V_{\rm kick}=0.6\,v_{\rm esc}$  
({\it bottom}). {\it Solid curves:} orbits in a NFW+bulge potential. {\it Dot-Dashed
curves:} orbits in a NFW potential. {\it Right panel:} Distributions of MBH displacements 
for all $z<3$ 
MBH binaries with $M_1+M_2>10^5\,\msun$ in our cosmological realizations. 
{\it Solid histogram}: nonspinning holes in a NFW potential.  
{\it Dotted histogram}: nonspinning holes in a NFW+bulge potential.  
{\it Dot-Dashed histogram}: $a=0.9$ holes in a NFW potential.  
{\it Dashed histogram}: $a=0.9$ holes in a NFW+bulge potential.  
}
\label{fig2}
\end{figure}
Following a kick, the radial orbit of a MBH in a spherical potential is governed by: 
\beq
{d^2 {\vec r} \over d t^2} =  -{G M(r) \over r^2}\, {\vec r} - 
{4 \pi G^2 \ln\Lambda\,\rho\, M_{\rm BH,tot} \over v^2} f(x)\,{\vec v}
\label{DF}
\eeq
where $f(x)\equiv [{\rm erf}(x) - (2 x/\sqrt{\pi}) e^{-x^2}]$, $x\equiv v/\sqrt{2}\sigma$, 
and the velocity dispersion $\sigma$ is derived from the Jeans' equation for the composite 
profile, assuming isotropy (e.g. \citealt{Binney1987}). Here $M(r)$ describes the total 
mass of the host galaxy within $r$, $\rho(r)$ is the total density profile, and the second 
term represents dynamical friction against the stellar and dark matter background 
(e.g. \citealt{Binney1987}). Stars and gas bound to the hole are displaced with it: here 
we assume, in first approximation (see below), that the total ejected mass 
is $M_{\rm BH,tot}=2M_{\rm BH}$. 
The Coulomb logarithm, $\ln \Lambda$, in equation (\ref{DF}) is taken equal to 2.5 
\citep{Gualandris2007}.

A MBH ejected with $V_{\rm kick}<v_{\rm esc}$ will oscillate about the nucleus 
losing orbital energy via dynamical friction \citep{MadauQuataert2004}. Note, however, 
that the assumption that the central stellar density profile is 
unmodified by the heating effect of dynamical friction tends to artificially 
shorten the MBH decay timescale \citep{Boylan-Kolchinetal2004}. Moreover, while 
in a spherical potential the hole 
will remain on a purely radial orbit, in a realistic triaxial galaxy the hole 
acquires angular momentum and does not return through the dense center. This is known to
reduce the effect of friction and to increase the decay time by a factor of a few 
relative to the 
spherical case \citep{Vicari2007}. To approximate the delaying effect of a triaxial geometry
and of a decreasing core density, we have also run a case in which MBHs move in 
in a bulgeless pure NFW potential. The two left panels in Figure \ref{fig2}.
show two examples of orbital decays (for $V_{\rm kick}=0.6,0.8\,v_{\rm esc}$) in a 
NFW+bulge (``short decay") and NFW only (``long decay'') model. The right panel depicts 
the ensuing distributions of MBH displacements for all $z<3$ kicked binaries 
in our cosmological realizations, with and without spin. In all but the NFW+bulge 
$a=0$ case, a significant fraction of recoiling holes are found between 10 and 
100 kpc from their galaxy centers (at redshift 2, a displacement of 10 kpc 
corresponds to an angular size of 1.2 arcseconds). Note that, in our realizations, 
only 0.1-0.3\% of all MBHs at a given cosmic epoch are actually off-center. 

\section{Off-nuclear AGN}

A recoiling hole that carries an accretion disk may be detected as it moves away 
from the center of its host. To assess the detectability of a subpopulation of 
off-nuclear AGN in deep imaging survey, we assume that all recoiling holes 
accrete at a fraction $f_E$ of the Eddington rate $\dot M_E=4\pi G M_{\rm BH}m_p/c\sigma_T$,
$L=\epsilon\,f_E\,L_E$, with radiative efficiency $\epsilon$. 
The rest-frame optical part of their spectrum is modelled as a multicolor 
\citet{ShakuraSunyaev} disk with maximum temperature $kT_{\rm max}\sim 
1\,{\rm keV}\,(M_{\rm BH}/\msun)^{-1/4}$, following a power-law with $L_\nu\sim 
\nu^{1/3}$ at $h\nu<kT_{\rm max}$. The X-ray spectral energy distribution is described 
by a power-law with photon index $\Gamma=1.9$ and 
an exponential cutoff at 500 keV (Marconi \etal 2004). The duration of the luminous phase 
depends on the amount of disk material out to the radius $R_{\rm out}\approx 
{GM_{\rm BH}/V_{\rm kick}^2}$ that is carried by the hole. In the case of an 
$\alpha$-disk, this is given by  \citep{Loeb2007}
\beq
M_{\rm disk}\approx (1.9\times 10^6\,\msun)~\alpha_{-1}^{-0.8} (\epsilon_{-1}
/f_E)^{-0.6} M_{7}^{2.2} V_{3}^{-2.8},
\label{mmax}
\eeq
where $\epsilon_{-1}\equiv \epsilon/0.1$, $M_{7}\equiv M_{\rm BH}/10^7\,\msun$,
$V_{3}\equiv V_{\rm kick}/10^3\,\kms$, and $\alpha_{-1} \equiv \alpha/0.1$ 
is the viscosity parameter. The condition $M_{\rm disk}\le M_{\rm BH}$ requires 
\begin{equation}
V_{\rm kick}\ge 550 \kms ~\alpha_{-1}^{-0.28} (\epsilon_{-1}/f_E)^{-0.21}M_7^{0.43}.
\label{vmax}
\end{equation}
When the above equation is not satisfied, we impose $M_{\rm disk}=M_{\rm BH}$ 
corresponding to an AGN lifetime of $t_{\rm max}=\epsilon c \sigma_T/(4\pi G 
m_p f_E)\approx 4.5\times 10^7\,{\rm yr}\,(\epsilon_{-1}/f_E)$. 
A hole/disk system with ($M_7,\alpha,\epsilon,f_E,M_{\rm disk})=(1,0.1,0.1,1,M_{\rm BH})$ 
recoiling with $V_{\rm kick}=500\,\kms$, could then still be shining as far as 30 kpc
away of the center of its host and appear as an off-center quasar.

\begin{figure}[thb]
\vspace{-0.3cm}
\includegraphics*[width=0.49\textwidth]{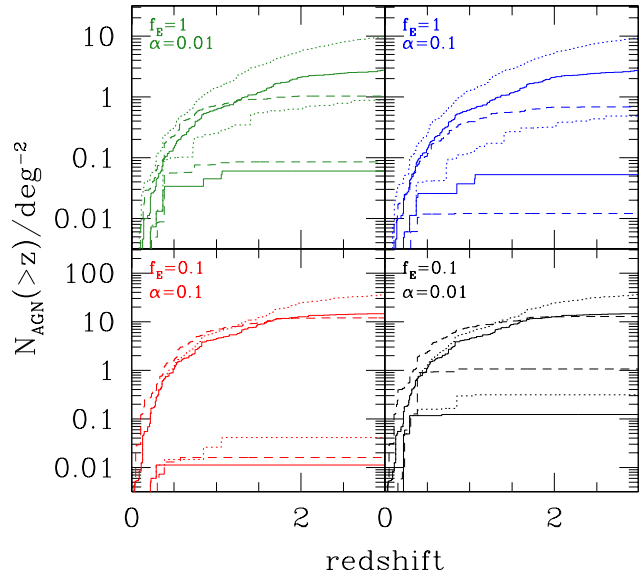}
\caption{\footnotesize Cumulative number counts of off-nuclear AGN detectable by the {\it HST}, 
{\it Chandra}, and the {\it JWST}. {\it Solid curve:} number counts at the sensitivity limit 
of the HST-COSMOS survey. {\it Dashed curve:} same at the sensitivity of the CDF-N survey. 
{\it Dotted curve}: same at a {it JWST} sensitivity of 10 nJy at 2$\mu$m. In each panel the 
upper (lower) set of curves shows the results for a bulgeless (NFW+bulge) potential. 
}
\label{fig3}
\end{figure}

Figure \ref{fig3} shows the expected number of detectable off-center AGN in deep
and large surveys such as the {\it Chandra} Deep Field North and {\it HST}-COSMOS, and the predicted 
counts for future observations with the {\it JWST}. We set the minimum resolvable separation 
between a recoiling MBH and its galaxy centroid to twice the 
telescope resolution (i.e. 0.2 arcsec for {\it HST} and {\it JWST}, 1 arcsec for {\it Chandra}). 
We do not correct our counts for obscuration or Compton thick sources. 
In the HST-COSMOS fields (2 square degrees) we expect $\sim 30$ sources for the best case of large 
kicks (spinning holes), long decay timescales (no bulge), and long active phase ($f_E$=0.1, $\alpha$=0.01), 
but less than 1 in the unfavourable cases. In the small CDF-N field we expect at most 1 
off-nuclear AGN in the most optimistic case, while in the 0.77deg$^2$ X--ray Chandra-COSMOS 
we can expect up to a few sources. Follow-up X--ray observations will be essential for the
identification of the source as a genuine off-center AGN. The {\it JWST} yields the highest counts,  about 40 sources per square degree in the best scenario.

\section{Summary}

Motivated by recent numerical simulations of black hole binary coalescence and ensuing 
gravitational-wave kicks, and by the large rate of MBH binary formation expected 
during the assembly of massive galaxies, we have assessed the detectability of recoiling
holes that retain the inner parts of their accretion disk along their trajectory and
shine as off-nuclear AGN. While the search for kinematically offset QSOs in the {\it Sloan 
Digital Sky Survey} (SDSS) indicates that large kicks rarely occur during a long-lasting active 
phase (with an upper limit on the incidence of wandering AGN with line-of-sight velocities 
above $800\,\kms$ of 0.2\%, \citealt{Bonningetal2007}), a strong candidate for a 
rapidly recoling MBH has been recently found in the SDSS database by \citet{Komossaetal2008}, 
showing an emission line spectrum offset by $2650\,\kms$ (but see Bogdanovic et al. 2008, Dotti et al. 2008 for a different interpretation). 
Here, we have traced the formation history of
MBH binaries in galaxies through cosmological Monte Carlo
realizations of the halo merger tree, assigned a kick velocity to every MBH binary coalescence
expected in our simulation, followed the orbital evolution of the ejected hole in the potential of the host, and determined the duration of a off-nuclear AGN phase on scales that can be
resolved by current and future optical, IR, and X-ray satellites. 
The basic conclusion of this Letter is that, depending on
the hole spin distribution and the duration of their active phase,
a population of off-nuclear AGN may already be detectable at low and intermediate redshifts in present deep {\it HST} observations. The {\it JWST} should discover tens of wandering AGN per square degree, moving within their host halos on unbound trajectories.

\acknowledgements{Support for this work was provided by NASA grants NNG04GK85G 
and NNX08AV68G (P.M.) and SAO-G07-8138 C (M.V.).}

\end{document}